\documentclass[twocolumn,preprintnumbers,amsmath,amssymb,showpacs,prb,aps]{revtex4}
\usepackage{graphicx}
\usepackage{dcolumn}
\usepackage{amsmath}
\usepackage[perpage,symbol*]{footmisc}

\makeatletter
\def\btt#1{\texttt{\@backslashchar#1}}%
\DeclareRobustCommand\bblash{\btt{\@backslashchar}}%
\makeatother

\begin{document}

\title{On the Munn-Silbey approach to polaron transport with off-diagonal coupling and temperature-dependent canonical transformations}

\author{Dongmeng Chen$^{1,2}$, Jun Ye$^{1}$, Haijun Zhang$^{1}$ and  Yang Zhao$^{1}$\footnote{Electronic address:~\url{YZhao@ntu.edu.sg}}}

\affiliation{\it $^1$ School of Materials Science and Engineering,
Nanyang Technological University, Singapore 639798\\
             \it $^2$ College
              of Physics Science and Technology,
             China University of Petroleum, Dongying 257061, China}
\date{\today}
\widetext

\begin{abstract}
Improved results using a method similar to the Munn-Silbey approach have been obtained on the temperature dependence of transport properties of an extended Holstein model incorporating simultaneous diagonal and off-diagonal exciton-phonon coupling. The Hamiltonian is partially diagonalized by a canonical transformation, and optimal transformation coefficients are determined in a self-consistent manner. Calculated transport properties exhibit substantial corrections on those obtained previously by Munn and Silbey for a wide range of temperatures thanks to a numerically exact evaluation and an added momentum-dependence of the transformation matrix. Results on the diffusion coefficient in the moderate and weak coupling regime show distinct band-like and hopping-like transport features as a function of temperature.
\end{abstract}

\maketitle \narrowtext

\section{Introduction}

Investigations on organic molecular crystals pioneered by Pope and cowokers \cite{Pope} more than half a century ago made an immense impact on various fields of relevance in physics, chemistry, and materials science. The interplay between geometric and electronic structures has opened up novel materials with unlimited possibilities. The discovery of conducting properties of doped polyacetylene \cite{Heeger} in 1977 led to a Nobel prize in chemistry establishing a whole new field of organic electronics. After decades of research, many families of the organic semiconducting materials have been investigated. Among them, oligoacenes \cite{Nelson,Lin,Dimi} and oligothiophenes \cite{Garnier} have been examined intensively due to the possibility of making single crystals with few defects via repeated sublimation processes. Single crystals of these materials can be used to develop transistors that allow for fundamental research on intrinsic properties as well as performance of organic electronic devices. The Holstein molecular crystal model was introduced in the 1950s \cite{Holstein1, Holstein2} to account for the effect of electron-phonon interactions on transport. Since then, a lot of work has been carried out to deal with the complexities of the transport phenomenon in organic molecular crystals. As reviewed in Ref.~\cite{ChemRev}, charge transport in organic semiconductors is governed by electronic coupling and electron-phonon interactions, and the main transport mechanism can be described by polaron and disorder models. Many theories on polaron transport are based upon a phenomenological model \cite{Silinsh} or utilize the polaron effective mass approach \cite{Berry}. However, validity of these models is restricted to specific parameter ranges. More general microscopic models in understanding polaron transport was developed \cite{Munn_Silbey1, Munn_Silbey2, Munn_Silbey3} from a density matrix approach, which is capable of describing electronic coupling, diagonal and off-diagonal electron-phonon interaction of arbitrary strength over a wide range of temperatures. The effect of off-diagonal coupling poses much more difficulties to tackle than its diagonal counterpart, and as a result, off-diagonal coupling has not been extensively treated before \cite{Mahan}. More recently, microscopic models from generalized master equation approach \cite{Kenkre1}, and dynamical mean-field theory \cite{DMFT1} have been developed.

Polaron transport theory developed by Holstein in his seminal work\cite{Holstein1, Holstein2} was based upon perturbation expansions with its validity sometimes limited to cases of small transfer integrals \cite{Holstein2}. Despite its limitation, the theory has been widely used for qualitative interpretations of experimental data, including the temperature dependence of band-narrowing effect as well as the crossover from band-like behavior to hopping transport with increasing temperature. A microscopic transport theory that accounts for simultaneous diagonal and off-diagonal electron-phonon coupling was developed by Munn and Silbey\cite{Munn_Silbey1, Munn_Silbey2, Munn_Silbey3} based on a canonical transformation of the Hamiltonian with an element of optimization. Off-diagonal coupling is found to increase the polaron binding energy and introduce new minima and broadening to the polaron band. In contrast, the effect of diagonal coupling is restricted to band narrowing. The Yarkony-Silbey variational approach\cite{YS1, YS2} provides an attractive direction to solve the transport problems in organic molecular crystals. This approach has been borrowed to treat two phonon bands in a three-dimensional lattice by Parris and Kenkre\cite{Kenkre2}. However, as the Yarkony-Silbey approach lacks the flexibility in variational parameters, a more sophisticated approach is needed to capture the correct temperature dependence of diffusion coefficient and mobility. To obtain the optimal basis for a polaron system, finite-temperature variational method by Cheng and Silbey \cite{Munn_Silbey4} combining Merrifield's transformation with Bogoliubov's theorem has been recently devised bringing substantial improvements to results obtained previously\cite{Munn_Silbey3}. The theory is able to capture the universal band-like-to-hopping transition as temperature increases, and more importantly, a temperature-independent mobility at extremely low temperature has been found in agreement with a nonperturbative approach recently developed by Ortmann \emph{et al.} \cite{Hannewald1, exp1}. However, Merrifield's transformation itself is applicable only in the small polaron regime, and the off-diagonal exciton-phonon coupling was not accounted for in Ref.~\cite{Munn_Silbey4}, making the approach insufficient to describe the phonon assisted transport/tunneling\cite{Munn_Silbey5} in molecular crystals. Thus, further improvements are necessary for cases with  simultaneous diagonal and off-diagonal coupling. Following up on the Munn-Silbey transformation method, Zhao \emph{et al.}\cite{zhaoMS, OD1,OD2} devised a self-consistent routine to determine the transformation coefficients, demonstrating substantial corrections of the polaron band structure thanks to a built-in momentum dependence of the transformation coefficients. This serves as the starting point towards the evaluation of transport properties in this paper.

An attempt to extend the Holstein model to higher dimensions with a microscopic model has been made by Kenkre \emph{et al.,}\cite{Kenkre1} utilizing the generalized master equation approach. This model was able to adequately explain the temperature dependence and anisotropy of measured mobilities in naphthalene. Using master equations, Wang \emph{et al.,} \cite{Wang} developed a nonperturbative method to handle the electron-phonon coupling fully quantum-mechanically and quantify charge-carrier transport properties in organic molecular crystals. A more sophisticated microscopic model based on a Hamiltonian of the Holstein-Peierls type has been presented by Bobbert and colleagues \cite{Hannewald2, Hannewald3, Hannewald4}. Very recently, a theory based on non-perturbative evaluation of Kubo formula for the carrier mobility\cite{Hannewald1, Hannewald5, Hannewald6} has been proposed, showing several improvements including the elimination of low-temperature singularity that often appears in theories based on narrow-band approximations. Based on a non-perturbative evaluation of the Kubo formula, calculations have been carried out for durene crystals\cite{Hannewald7} and guanine based materials\cite{Hannewald8}, and transport channels of charge carriers are  revealed by making use of ab-initio tools. Coupling to the intermolecular acoustic modes \cite{Kenkre3} was found to play a significant role in charge transport, thus the model presented by Bobbert \emph{et al.} may be improved further. Experimental data of naphthalene can be well reproduced by this model with microscopic parameters obtained from ab-initio calculations. However, this model neglects the intra-molecular modes as well as the coupling of excitons to the intermolecular acoustic phonons, and only the coupling to the optical modes is accounted for. Moreover, the models developed by Bobbert \emph{et al.,} and by Munn and Silbey, are based on nonlocal canonical transformations with additional approximations. Thus the range of validity of these models still requires further exploration, despite some qualitative agreements with experiments. Hence, in the light of recent theories by Hultell \emph{et al.}\cite{Hultell} and Troisi \emph{et al.}\cite{Troisi}, it is clear that both local and nonlocal exciton-phonon interactions should be taken into account. A comprehensive theory of the charge transport in organic crystals requires treatments of the Hamiltonian having the lattice dealt with quantum mechanically in order to make itself valid for any temperature of interest, and the Munn-Silbey approach\cite{Munn_Silbey1, Munn_Silbey2, Munn_Silbey3} combined with the self-consistent routine devised by Zhao \emph{et al.}\cite{zhaoMS} provides a good opportunity to explore the simultaneous effect of diagonal and off-diagonal exciton-phonon coupling on the transport properties of polaron.

The rest of the paper is organized as follows. The theoretical models are firstly introduced in Section II, where the essential improvements over the original Munn-Silbey model have been proposed. In Section III, the numerical results obtained with our approach have been discussed in detail and comparisons with previous results are made to illustrate the success of our approach. Finally, the conclusions are drawn in section IV.

\section{Methodology}

The generalized Holstein Hamiltonian and its transformation have been described in detail previously by Munn and Silbey\cite{Munn_Silbey1, Munn_Silbey2, Munn_Silbey3}. Here we will first discuss briefly the canonical transformation of the Hamiltonian which yields a weak residual excitation-phonon coupling. Evaluation of correlation functions, essential to the final calculation of diffusion coefficient and mobility as a function of temperature, follows next. Our major focus in this paper will be the explicit form of the correlation functions, evaluation of critical exponential matrices and especially the transformation coefficients without resorting to the approximations in Ref.~\cite{Munn_Silbey2}. In our study, the transformation coefficients are determined through solving a set of self-consistency equations, rather than an analytical form approximated in Ref.~\cite{Munn_Silbey2}, therefore providing substantial corrections in the transport calculations. In Sec.~II A and B, expressions for the scattering and the hopping rates as well as the diffusion coefficients are introduced. Correlation functions are presented in Sec.~II C.

\subsection{Hamiltonian and its Transformation}

The generalized Holstein Hamiltonian incorporating simultaneous
diagonal and off-diagonal coupling \cite{Munn_Silbey2} has the form:
\begin{eqnarray}\label{Hamiltonian0}
H&=&\sum_{n}{\epsilon}a_{n}^{\dagger}a_{n} +\sum_{n,m}J_{nm}a_{n}^{\dagger}a_{m} \nonumber \\
&& +\sum_{q}\omega_{q}(b_{q}^{\dagger}b_{q}+1/2) \nonumber \\
&& +N^{-1/2}\sum_{nmq}\omega_{q}f_{nm}^{q}a_{n}^{\dagger}a_{m}(b_{q}+b_{-q}^{\dagger}),
\end{eqnarray}
here $a_{n}^{\dagger}(a_{n})$ is the creation (annihilation) operator of an excitation (i.e., exciton or charge carrier) with energy $\epsilon$ at site n, and $b_{q}^{\dagger}(b_{q})$ creates (destroys) a phonon with frequency $\omega_q$ and wave vector \emph{q}. $J_{nm}$ stands for the electronic transfer integral between site $n$ and $m$. In this study, the discrepancies between molecules have been neglected, thus the on-site energies are set to $\epsilon$ as shown in Eq.\ref{Hamiltonian0}. Finally, the excitation-phonon coupling is described by the last term of Eq.\ref{Hamiltonian0}, where the coupling strength $f_{nm}^{q}$ must satisfy the relationship : $(f_{n-m}^{q})^* = e^{-i\textbf{q}\cdot(\textbf{R}_n - \textbf{R}_m)}f_{n-m}^{-q}$, to make $H$ Hermitian and translationally invariant.

A unitary transformation can be applied to the Hamiltonian, as shown below:
\begin{equation}\label{transform1}
H\rightarrow\tilde{H}=U^{\dagger}HU,
\end{equation}
with
\begin{equation}\label{transform2}
U=e^{N^{-1/2}\sum_{nmq}A_{nm}^{q}(b_{-q}^{\dagger}-b_{q})a_{n}^{\dagger}a_{m}}.
\end{equation}

The transformation requires $U^{\dagger}=-U$ in order to keep the translational symmetry of $A_{nm}^q$, similar to $f_{nm}^{q}$. Details of the transformation have been explicitly introduced in Ref.~\cite{Munn_Silbey1, Munn_Silbey2, Munn_Silbey3}.

Moreover, the Hamiltonian in momentum representation is more convenient than the site representation due to the translational symmetry. Eq.~\ref{Hamiltonian0} can be written in the momentum representation as:
\begin{eqnarray}\label{Hamiltonian1}
H&=&\sum_{k}{\epsilon_{k}}a_{k}^{\dagger}a_{k} + \sum_{q}\omega_{q}(b_{q}^{\dagger}b_{q}+1/2) \nonumber \\
&& +N^{-1/2}\sum_{kq}\omega_{q}f_{-k}^{q}a_{k+q}^{\dagger}a_{k}(b_{q}+b_{-q}^{\dagger}),
\end{eqnarray}
with $\epsilon_{k}=\epsilon+J_{k}$, and
\begin{equation}\label{jk}
J_{k}=\sum_{m}e^{ik\cdot(\textbf{R}_{n}-\textbf{R}_{m})}J_{nm},
\end{equation}
where $J_{n,m}=J(\delta_{n,m+1}+\delta_{n,m-1})$, and $\omega_{q}$ is the phonon frequencies.

The excitation-phonon coupling coefficients can also be written as:
\begin{equation}\label{fk}
f_{k}^{q}=\sum_{m}e^{-ik\cdot(\textbf{R}_{n}-\textbf{R}_{m})}f_{n-m}^{q},
\end{equation}
with following coupling geometry:
\begin{equation}\label{couplinggeometry}
f_{k}^{q}=g-i\phi[{\rm sin}(k)-{\rm sin}(k-q)],
\end{equation}
which indicates that the antisymmetric type of coupling is used throughout the paper.

Now the transformation takes the following form in the momentum representation:
\begin{equation}\label{uk}
U=e^{N^{-1/2}\sum_{kq}A_{-k}^{q}(b_{-q}^{\dagger}-b_{q})a_{k+q}^{\dagger}a_{k}},
\end{equation}
and
\begin{equation}\label{ak}
A_{k}^{q}=\sum_{m}e^{-i(\textbf{k}-\textbf{q})\cdot\textbf{R}_{n}}e^{i\textbf{k}\cdot\textbf{R}_{m}}A_{nm}^{q}=(A_{k-q}^{-q})^*.
\end{equation}

The transformation can then also be written as:
\begin{equation}\label{uk2}
U=e^{\sum_{kk'}a_{k}^{\dagger}\textbf{S}_{kk'}a_{k'}},
\end{equation}
with
\begin{equation}\label{sk}
\textbf{S}_{kk'}=N^{-1/2}A_{-k}^{k-k'}(b_{k'-k}^{\dagger}-b_{k-k'}).
\end{equation}
where $\textbf{S}_{kk'}$ is an operator creating a net phonon with momentum $k'-k$. This operator is essential towards the evaluation of a series of important matrices, where
\begin{eqnarray}\label{ak}
a_k\rightarrow\sum_{k'}\theta_{kk'}a_{k'} \nonumber \\
a_{k}^{\dagger}\rightarrow\sum_{k'}\theta_{kk'}^{\dagger}a_{k'}^{\dagger}
\end{eqnarray}
with
\begin{eqnarray}\label{thetak}
\theta_{kk'}=[{\rm exp}(-\textbf{S})]_{kk'} \nonumber \\
\theta_{kk'}^{\dagger}=[{\rm exp}(\textbf{S})]_{kk'}
\end{eqnarray}
where $\textbf{S}$ stands for matrix ${\textbf{S}_{kk'}}$. The transformed Hamiltonian can be partitioned into $H_0$ terms (zeroth order part) and the perturbative $V$ terms. Thermal average of the $\theta_{kk'}$ operator has the property:
\begin{equation}\label{propertytheta}
\langle\theta_{k,k+q}^{\dagger}\theta_{k',k'+q'}\rangle=\langle\theta_{k,k+q}^{\dagger}\theta_{k',k'+q}\rangle\delta_{qq'},
\end{equation}
which ensures the correct thermal equilibrium behavior of $H_{0}$ by keeping it diagonal in excitation wave vector. $H_{0}$ here has following expression:
\begin{eqnarray}\label{Hamiltonian2}
\tilde{H}_0&=&\sum_{q}\omega_{q}(b_{q}^{\dagger}b_{q}+1/2)+\sum_{k}a_{k}^{\dagger}a_{k}\{(\epsilon-\tilde{J}_k)\nonumber\\
&+& N^{-1}\sum_{q}\omega_{q}|A_{-k}^{-q}|^2-2N^{-1}\nonumber \\
&\times&\sum_{qk'}\omega_{q}f_{-k'}^{q}A_{-k}^{-q}\langle\theta_{k'+q,k}^{\dagger}\theta_{k',k-q}\rangle +N^{-1/2}\nonumber \\
&\times&\sum_{qk'}\omega_{q}f_{-k'}^{q}A_{-k}^{-q}\langle\theta_{k'+q,k}^{\dagger}\theta_{k',k}(b_{q}+b_{-q}^{\dagger})\rangle\},
\end{eqnarray}
here $\tilde{J}_k$ represents the renormalized transfer integral which can be expressed by:
\begin{equation}\label{propertytheta}
\tilde{J}_k=\sum_{k'}J_{k'}\langle\theta_{k'k}^{\dagger}\theta_{k'k}\rangle.
\end{equation}

The perturbation parts of the transformed Hamiltonian can be written as:
\begin{eqnarray}\label{Hamiltonian3}
V&=&\sum_{kk'k''}\{J_{k}T_{kk';kk''}-2N^{-1}\sum_{q}\omega_{q}f_{-k}^{-q}(A_{-k''}^{q})\nonumber\\
&\times&{T_{k+q,k';k,k''-q}}+N^{-1/2}\sum_{q}\omega_{q}f_{-k}^{q}a_{k'}^{\dagger}a_{k''}\nonumber\\
&\times&[T_{k+q,k';kk''}(b_q+b_{-q}^{\dagger})-\langle{T_{k'+q,k';kk''}}(b_{q}+b_{-q}^{\dagger})\rangle]\}\nonumber\\
&+&N^{-1/2}\sum_{qk}\omega_{q}[A_{-k}^{q}-\sum_{k'}f_{-k'}^{q}\langle\theta_{k'+q,k}^{\dagger}\theta_{k',k-q}\rangle]\nonumber\\
&\times&(b_{q}+b_{-q}^{\dagger})a_{k+q}^{\dagger}a_k,
\end{eqnarray}
where
\begin{equation}\label{propertytheta}
T_{kk';uu'}=\theta_{kk'}^{\dagger}\theta_{uu'}\ - \langle\theta_{kk'}^{\dagger}\theta_{uu'}\rangle.
\end{equation}

However, the last term of $V$ has the potential to grow with increasing temperature, motivating the choice of $A_k^q$:
\begin{equation}\label{secular}
A_{-k}^{q}=\sum_{k'}f_{-k'}^{q}\langle\theta_{k'+q,k}^{\dagger}\theta_{k',k-q}\rangle,
\end{equation}
to curb this possible uncontrolled growth of $\theta_{kk'}$.

By applying the thermal average routine shown in Ref.\cite{Munn_Silbey2}, the transformed Hamiltonian can be written in a simpler form, which facilitates the numerical calculations. The zeroth order Hamiltonian $H_0$ now can be written as:
\begin{equation}\label{Hamiltonian4}
\tilde{H}_0=\sum_{k}\tilde{\epsilon}_{k}a_{k}a_{k}+\sum_{q}\omega_{q}(b_{q}^{\dagger}b_{q}+\frac{1}{2}),
\end{equation}
and
\begin{equation}\label{epsilontilde}
\tilde{\epsilon}_{k}=\epsilon+\tilde{J}_{k}-N^{-1}\sum_{q}|A_{k}^{q}|^2\omega_{q}.
\end{equation}
The renormalized transfer integral in Eq.~\ref{epsilontilde} is given by:
\begin{equation}\label{Jtilde}
\tilde{J}_{k}=\sum_{k'}{\langle\theta_{k}\rangle}^2({\rm exp}E^0)_{kk'}J_{k'},
\end{equation}
where the triadic $E_{kk'}^{q}$ is a quantity introduced to yield a simpler form of the transformed Hamiltonian, with following relationship:
\begin{equation}\label{selfconsistent1}
E_{kk'}^{q}=N^{-1}(2n_{k-k'}+1){(A_{k-q}^{k-k'})^*}{A_{k}^{k-k'}},
\end{equation}
here $n_q$ is the Bose-Einstein distribution, with $2n_q+1={\rm coth}(\frac{1}{2}\beta\hbar\omega)$.

With the form of $A_{k}^{q}$ shown in Eq.~\ref{secular}, the perturbation part $V$ of $\tilde{H}$ can be written as:
\begin{eqnarray}\label{perturbation2}
V&=&\sum_{kk'k''}\{J_{k}T_{kk';kk''}-2N^{-1} \nonumber\\
&\times&\sum_{q}\omega_{q}f_{-k}^{q}A_{-k''}^{-q}{T}_{k+q,k';k,k''-q}+N^{-1/2}\nonumber\\
&\times&\sum_{q}\omega_{q}f_{-k}^{q}T_{k+q,k';kk''}(b_q+b_{-q}^{\dagger})a_{k'}^{\dagger}a_{k''}\}.
\end{eqnarray}

In this study, the transformation coefficients $A_{k}^{q}$ (A-matrix) will be obtained variationally prior to the calculations of the transport properties. The detailed numerical variational procedures can be found in Ref.~\cite{zhaoMS}. In performing the variational calculations, the transformation coefficients must satisfy following self-consistency equations together with Eq.~\ref{selfconsistent1}:
\begin{equation}\label{selfconsistent2}
\langle\theta_{k}\rangle={\rm exp}[-\frac{1}{2}\sum_{k'}E_{kk'}^{0}],
\end{equation}
\begin{equation}\label{selfconsistent3}
A_{k}^{q}=\langle\theta_{k-q}\rangle\langle\theta_{k}\rangle\sum_{k'}f_{k'}^{q}[{\rm exp}(E^{q})]_{kk'}.
\end{equation}

The self-consistency equations above follow the Munn-Silbey secular elimination scheme. In Ref.~\cite{Munn_Silbey2}, this set of equations was solved upon further  approximates on $A_k^q$, whereas in this paper, it is solved numerically with an accuracy unavailable previously.

In the Munn-Silbey approach\cite{Munn_Silbey2}, the transformation coefficients $A_{k}^q$ was at first approximated by the scaling parameters $\xi$ and $\eta$ as:
\begin{equation}\label{variational1}
A_{k}^{q}=g\xi-i\phi\eta[{\rm sin}(k)-{\rm sin}(k-q)].
\end{equation}

Numerical evaluation of the Eqs.~\ref{selfconsistent1}-\ref{selfconsistent3} is facilitated by rewriting the Eq.~\ref{variational1} with real matrices $\xi_{k}^{q}$ and $\eta_{k}^{q}$ in the following form:
\begin{equation}\label{variational2}
A_{k}^{q}=g\xi_{k}^{q}-i\phi\eta_{k}^{q}[{\rm sin}(k)-{\rm sin}(k-q)].
\end{equation}

In the equation above, the sine function leads to zero imaginary part along the line with $q=0, 2k\pm{\pi}$. In the numerical calculation, the lines with $q=0$ and $q=2k\pm{\pi}$ are treated as removable singularities. Moreover, the values of $\eta_{k}^{q}$ along these lines are chosen to be analytically connect with neighboring values\cite{zhaoMS}. The variational matrices $\xi_{k}^{q}$ and $\eta_{k}^{q}$ chosen can preserve the symmetry properties of $A_{k}^{q}$ in a way such that $\xi_{k}^{q}=\xi_{k-q}^{-q}$ and $\eta_{k}^{q}=\eta_{k-q}^{-q}$.

The self-consistency approach shown in this paper is largely limited by the secular elimination scheme that is independent on the rigid-lattice tunneling matrix elements $``J"$. However, since $``J"$ determines the rigid-lattice band structure and related quantities such as effective mass and density of states, this approach only applies for the narrow (rigid) band regime where the traditional small polaron theory is still available. The wide (rigid) band systems are beyond the scope of the approach discussed in this work. Based on the simple introduction and discussion of the Hamiltonian and transformation schemes, the parameters used for our numerical calculations are limited to the regime which holds the traditional small polaron scenario.

\subsection{Diffusion Coefficient}

The diffusion coefficient is closely related to the mean square displacement of an excitation in a given period of time. For the transformation scheme shown in this paper, the mean square displacement can be calculated from the excitation density matrix. Approximations have been made to reduce the complexity of the exact formulism of the mean diffusion coefficient by neglecting small terms of the perturbation $V$ as shown in the section above. The diffusion coefficient can then be expressed as\cite{Munn_Silbey1,Munn_Silbey3}:
\begin{equation}\label{DD1}
D=a^{2}\langle\langle\nu_{k}^{2}/\Gamma_{kk}+\gamma_{kk}\rangle\rangle,
\end{equation}
where $a$ is the nearest neighbor distance, $\textbf{k}$ is the wave vector, and $\nu_k=\nabla_{k}E_{k}$. The double bracket denotes the thermal average of the polaron states $E_k$, where $\Gamma_{kk}$ and $\gamma_{kk}$ are the scattering and hopping rates expressed as
\begin{eqnarray}\label{scattering}
\Gamma_{k'k'}& =&N^{-1}\sum_{k{\neq}k'}W_{k,k;k',k'},\\
\gamma_{kk}&=&\frac{1}{2}\nabla_{q}^2\sum_{k}\nonumber \\
&{\rm Re}& (\frac{1}{2}W_{k,k;k'+q,k'+q}-W_{k,k+q;k',k'+q})|_{q=0}.
\end{eqnarray}
where $W_{k,k+q;k',k'+q}$, crucial to the calculation of diffusion coefficient, are given by
\begin{eqnarray}\label{scattering}
&&W_{k,k+q;k',k'+q}\nonumber \\
&=&\int_{0}^{\infty}dt\{\langle{V_{k'+q,k+q}V_{k,k'}(t)}\rangle{\rm exp}[-i(E_{k'+q}-E_{k+q})t] \nonumber \\
&+&\langle{V_{k'+q,k+q}(t)V_{kk'}}\rangle{\rm exp}[-i(E_{k}-E_{k'})t]\},
\end{eqnarray}
here the single angular bracket denotes the average of the residual interaction $V_{kk'}(t)$ over phonon states. Using the formulism above, the evaluation of $D$ will be done numerically here without introducing further approximations as in Ref.~\cite{Munn_Silbey1,Munn_Silbey2}. With the diffusion coefficient D, the mobility can be easily obtained with the Einstein relation $\mu=e\beta{D}$.

\subsection{Correlation Functions}

As inferred from the expression of the diffusion coefficient, the central problem is the evaluation of the correlation functions $\langle{V_{k'+q,k+q}V_{k,k'}(t)}\rangle$ since both the scattering and the hopping rates are determined by the $W$ quantities. It is indicated from the previous formulism that the residual interaction $V_{k,k'}(t)$ is closely related to the transformation coefficients $A_{k}^{q}$. Hence, the problem of evaluating $D$ ultimately hinges on the solution of the self-consistency equations, i.e., Eqs.~\ref{selfconsistent1}, \ref{selfconsistent2} and \ref{selfconsistent3}. In this paper, the transformation coefficients that determine the structure of polaron states and energy bands are solved numerically, which provide new information about the transport properties of polarons in molecular crystals.

Without the analytical approximation form of $\eta$ carried out in Ref.~\cite{Munn_Silbey3}, the important quantity $\langle\theta _{k,k^{\prime }}^{\dag }\theta _{q,q^{\prime }}\rangle$ can be derived as\cite{Munn_Silbey2} :
\begin{eqnarray}\label{twotheta}
\langle \theta _{k,k^{\prime }}^{\dag }\theta _{q,q^{\prime }}\rangle
&=&\langle \theta _{-k^{\prime }}\rangle \exp (D^{k-q})_{k,k^{\prime}}\langle \theta _{-q^{\prime }}\rangle \delta _{k-k^{\prime },q-q^{\prime }} \nonumber \\
&=&\langle \theta _{-k^{\prime }}\rangle \exp (E^{k-q})_{-q^{\prime},-q}\langle \theta _{-q^{\prime }}\rangle \delta _{k-k^{\prime},q-q^{\prime }}, \nonumber \\
\end{eqnarray}
and
\begin{eqnarray}
(E^{q})_{k,k^{\prime }}[t] &=&\frac{1}{N}P_{k-k^{\prime }}(t)(A_{k-q}^{k-k^{\prime }})^{\ast }A_{k}^{k-k^{\prime }}, \\
(D^{q})_{k,k^{\prime }}[t] &=&\frac{1}{N}P_{k-k^{\prime }}(t)(A_{-k^{\prime }}^{k-k^{\prime }})^{\ast }A_{-k^{\prime }+q}^{k-k^{\prime }},
\end{eqnarray}
where the expression of $\langle\theta_{k}\rangle$ is given by Eq.~\ref{selfconsistent2}.

The factors $P_Q(t)$ are the characteristics of correlation functions for linear exciton-phonon coupling \cite{Grover_Silbey}, which may take the form
\begin{eqnarray}\label{P_t}
P_Q(0)&=& (2n_Q+1)/N\nonumber\\
P_Q(t) &=& [(2n_Q+1){\rm cos}(\omega{t})+i{\rm sin}(\omega t)]/N,
\end{eqnarray}
where $n_Q$ is also the Bose-Einstein distribution, with $(2n_Q+1)={\rm coth}(\frac{1}{2}\beta\hbar\omega)$ and $\omega$ is the mean phonon frequency.

In order for the correlation to decay to zero at long times, $P_Q(t)$ should be multiplied by a decay factor $A(t)$, defined as the Fourier transform of the phonon density of states, which takes a Gaussian form \cite{Munn_Silbey1} as given by:
\begin{equation}\label{phononDOS}
A(t)={\rm exp}(-\Delta^2{t}^2/4),
\end{equation}
where $\Delta$ is the phonon bandwidth taken much less than $\omega$. For the convenience of numerical calculations, the mean phonon frequency $\omega$ is set to 1. For all calculations performed in this paper, $\Delta$ is set to be 0.1, indicating a rather narrow phonon band\cite{Munn_Silbey1}. The inclusion of phonon bandwidth is a natural result of exciton-phonon coupling. For the Holstein polaron model, it has been inferred \cite{Holstein2, Emin1} that phonon dispersion is directly linked to lattice relaxation time, and it is also crucial for the calculation of probabilities of hopping events accompanied simultaneous absorption and emission of phonon. It is also pointed out by Holstein\cite{Holstein2} that replacing the phonon dispersion by a single frequency (Einstein model) yields meaningless results for probabilities of hopping events. A number of factors can result in a finite bandwidth in the phonon density of states, and as pointed out by other authors \cite{Munn_Silbey1,Munn_Silbey4}, the inclusion of the phonon bandwidth ensures the long-time decay of correlation functions.

In the evaluation of $\langle{V_{k'+q,k+q}V_{k,k'}(t)}\rangle$, an important quantity is the four $\theta_k$ correlation function which is given by:
\begin{eqnarray}\label{fourtheta}
&&\langle \theta _{k_1,k_2}^{\dag }\theta _{k_3,k_4}\theta _{q_1,q_2}^{\dag}(t)\theta _{q_3,q_4}(t)\rangle \nonumber \\
&=&\underset{r_1,r_2,r_3}{\sum }\langle \theta_{k_2+r_1,k_2}^{\dag }\theta _{k_3,k_3-r_1}\rangle \langle \theta_{q_2+r_2,q_2}^{\dag }\theta _{q_3,q_3-r_2}\rangle \nonumber \\
&\times&[\exp (-E^{k_1+q_4+r_3-q_2-q_3}(t))]_{k_3+r_3-q_1-k_4-r_1,-q_2-r_2} \nonumber \\
&\times&[\exp (E^{k_1+r_2-q_3}(t))]_{-q_4-r_3,r_2-q_3} \nonumber \\
&\times&[\exp (E^{k_3-q_1-r_1}(t))]_{-q_1,k_3+r_3-q_1-k_4-r_1} \nonumber \\
&\times&[\exp (-E^{k_4-q_4-r_3}(t))]_{-q_4,-q_4-r_3} \nonumber \\
&\times&\delta _{k_1+q_1+k_4+q_4,k_2+q_2+k_3+q_3}.
\end{eqnarray}

Now we can derive the expression for $\langle{V_{k'+q,k+q}V_{k,k'}(t)}\rangle$, which is the core item of W. Here we first express $V_{k,k'}$ as follows:
\begin{eqnarray*}\label{Vkk}
V_{k,k^{\prime }} &=&\underset{\kappa }{\sum }\{J_{\kappa }T_{\kappa k,\kappa k^{\prime }}-\frac{1}{\sqrt{N}}\underset{q}{\sum }\omega _{q}f_{-\kappa}^{q} \\
&\times&[\frac{1}{\sqrt{N}}(A_{-k}^{-q}T_{\kappa +q,k;\kappa ,k^{\prime}-q}+A_{-k}^{q\ast }T_{\kappa +q,k+q;\kappa ,k^{\prime }}) \\
&-&\frac{1}{2}(T_{\kappa +q,k;\kappa ,k^{\prime }}\psi _{q}+\psi_{q}T_{\kappa +q,k;\kappa ,k^{\prime }})]\}
\end{eqnarray*}
with
\begin{equation}
T_{k,k^{\prime };q,q^{\prime }}=\theta _{k,k^{\prime }}^{\dag }\theta_{q,q^{\prime }}-\langle \theta _{k,k^{\prime }}^{\dag }\theta _{q,q^{\prime}}\rangle,
\end{equation}
and $J_{k}=J\cos (k)$ denotes the polaron band with its width determined by $J$. As described in the previous section, the currently applied approach is only applicable for narrow band rigid lattice system. Consequently, the value of $J$ should not be too large in order to guarantee the accuracy of the transport results obtained.

The correlation function is now given by:
\begin{eqnarray}\label{VVT1}
&&\langle V_{k_1,k_2}V_{q_1,q_2}(t)\rangle \nonumber\\
&=&\langle \underset{k_3}{\sum }\{J_{k_3}T_{k_3,k_1;k_3,k_2}-\frac{1}{\sqrt{N}}\underset{Q_1}{\sum }\omega_{Q_1}f_{-k_3}^{Q_1} \nonumber\\
&\times&[\frac{1}{\sqrt{N}}(A_{-k_1}^{-Q_1}T_{k_3+Q_1,k_1;k_3,k_2-Q_1}\nonumber\\
&+&A_{-k_1}^{Q_1\ast}T_{k_3+Q_1,k_1+Q_1;k_3,k_2})\nonumber\\
&-&\frac{1}{2}(T_{k_3+Q_1,k_1;k_3,k_2}\psi _{Q_1}+\psi_{Q_1}T_{k_3+Q_1,k_1;k_3,k_2})]\} \nonumber\\
&\times&\underset{k_4}{\sum }\{J_{k_4}T_{k_4,q_1;k_4,q_2}-\frac{1}{\sqrt{N}}\underset{Q_2}{\sum }\omega _{Q_2}f_{-k_4}^{Q_2} \nonumber\\
&\times&[\frac{1}{\sqrt{N}}(A_{-q_1}^{-Q_2}T_{k_4+Q_2,q_1;k_4,q_2-Q_2}\nonumber\\
&+&A_{-q_1}^{Q_2\ast}T_{k_4+Q_2,q_1+Q_2;k_4,q_2})\nonumber\\
&-&\frac{1}{2}(T_{k_4+Q_2,q_1;k_4,q_2}\psi _{Q_2}+\psi_{Q_2}T_{k_4+Q_2,q_1;k_4,q_2})]\}\rangle\nonumber. \\
\end{eqnarray}
The explicit expression of the correlation function has been given in the Appendix A.

As read from the expression of the correlation function in Eq.~\ref{VVT1}, matrices $A_k^{q}$ play an essential role towards an accurate evaluation of the correlation function. In Ref.~\cite{Munn_Silbey3}, these matrices were obtained only approximately, while in this paper, they are obtained numerically with unprecedented accuracy. One will find that the accurate evaluation of the $A_k^{q}$ has dramatic influence on the overall transport calculations in the following section, especially for the cases with off-diagonal excitation-phonon coupling.

\section{Results and Discussion}

\subsection{Evaluation of Transformation Coefficients}

\begin{figure}[tb]
\begin{center}
\includegraphics[scale=0.8]{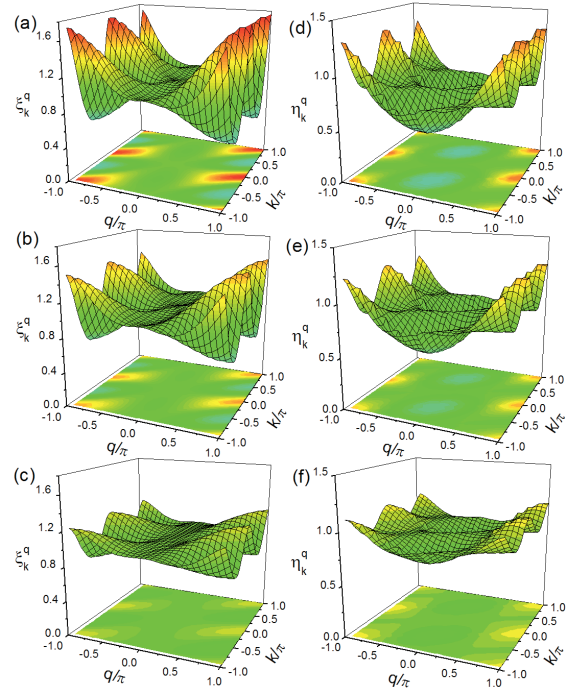}
\end{center}
\caption{A-matrix parameters $\xi_k^{q}$ (a)-(c) and $\eta_{k}^{q}$ (d)-(f) as functions of increasing diagonal coupling strength $g^2$. $g^2=0.1$ for (a) and (d), $0.3$ for (b) and (e), and $1.0$ for (c) and (f). The other parameters are: J=0.1, $\phi^2$=0.3 and T=1.0.}\label{figure1}
\end{figure}

The numerical evaluation of transformation coefficients $A_k^{q}$ involves the solution of the self-consistency equations Eqs.~\ref{selfconsistent1}, \ref{selfconsistent2} and \ref{selfconsistent3}. Details of the evaluation procedures have been discussed in detail in Ref.~\cite{zhaoMS}, here we only revisit some important issues related to the transport properties, especially the role of off-diagonal coupling.

For moderate temperature with $k_{\rm B}T=\hbar\omega$, the self-consistency method applied in this paper tends to fail in strong coupling regimes, this is attributed to the failure to minimize errors to desired precision or converging to multiple solutions that are sensitive to procedure initialization. It is also possible for the convergence to fail in strong coupling regimes for solely numerical reasons. This problem increases with increasing excitation-phonon coupling, which lead to large positive and negative eigenvalues for the exponential matrices, thus to the possible failure of obtaining reliable diffusion coefficient. Due to the problems of convergence and possible multiple solutions of the method applied, we restricted our studies in the weak and intermediate coupling regimes to ensure the robustness of the results obtained.

As demonstrated in Ref.~\cite{zhaoMS}, A-matrix parameters $\xi_{k}^{q}$ and $\eta_{k}^{q}$ in Eq.~\ref{variational2} obtained with the self-consistency procedure have a significant wave vector dependence, which is lacking in the original treatment of Munn and Silbey \cite{Munn_Silbey2} as indicated in Eq.~\ref{variational1}. In addition, the Munn-Silbey parametrization of the A matrix, $\eta$ and $\xi$, can be regarded as the lower bounds of $\xi_{k}^{q}$ and $\eta_{k}^{q}$ rather than the average values.  More importantly, the wave vector dependence of the variational parameters lead to significant changes in the structure of polaron energy band $E_k$, where the bimodal variation of $E_k$ can be always found. Such changes in the structure of polaron states and the energy band lead to critical modifications of the transport properties, which will be explicitly examined in the next section. In this section, we present detailed structures of A-matrix parameters as well as the corresponding polaron bands obtained with the self-consistency routines, in order to facilitate understanding of calculated transport properties, which are deferred to the next subsection.

\begin{figure}[tb]
\begin{center}
\includegraphics[scale=0.36]{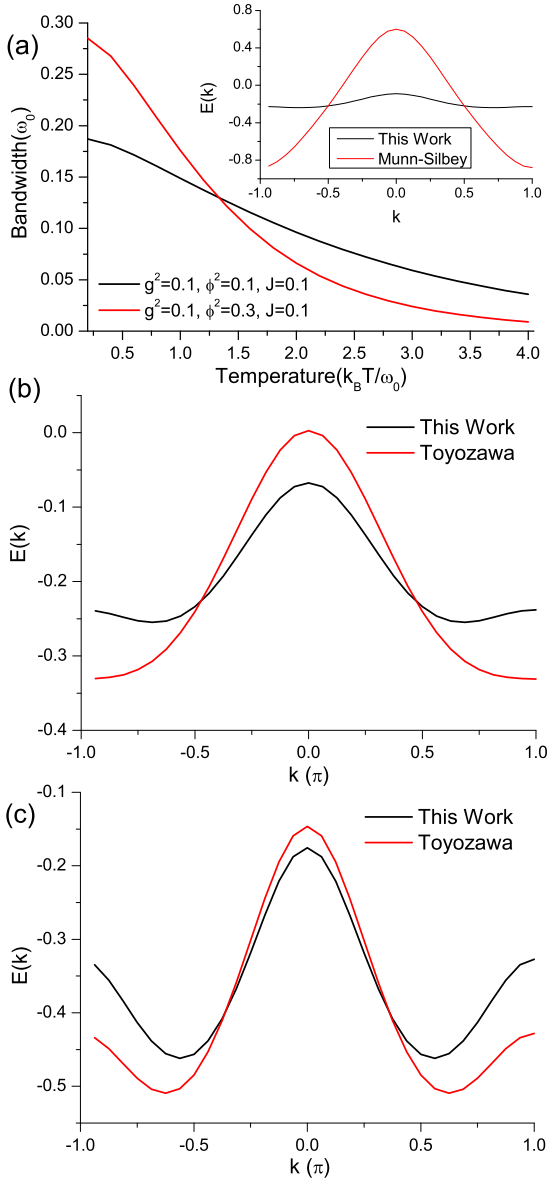}
\end{center}
\caption{Polaron bandwidths (a) as functions of temperature, and band structures (b)-(c) comparisons between this work and those from Toyozawa variational method with off-diagonal coupling strength (b) $\phi^2$=0.1 and (c) $\phi^2$=0.3 at T=0, where $g^2$ and $J$ are both equal to 0.1 for all calculations. The inset of (a) is the band structure comparison between the result from Ref.~\cite{Munn_Silbey2} by Munn and Silbey and that from this work at T=1 with $g^2$ = $\phi^2$ = $J$ = 0.1.}\label{figure2}
\end{figure}

\begin{figure}[tb]
\begin{center}
\includegraphics[scale=0.8]{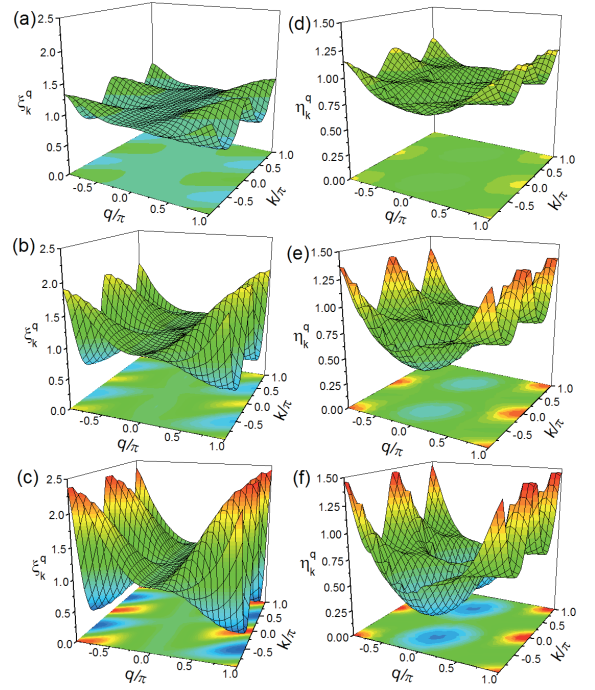}
\end{center}
\caption{A-matrix parameters $\xi_k^{q}$ (a)-(c) and $\eta_{k}^{q}$ (d)-(f) as functions of increasing temperature.
$ T = 0.2$ for (a) and (d) , $1.3$ for (b) and (e), and $4.0$ for (c) and (f). Other parameters are J=0.1, $g^2$=0.1 and $\phi^2$=0.3.}\label{figure3}
\end{figure}

As shown in Fig.~\ref{figure1}, the two sets of A-matrix parameters in Eq.~\ref{variational2}, $\xi_{k}^{q}$ and $\eta_{k}^{q}$, differ in their wave vector dependence. In the vicinity of $q=0$, $\xi_{k}^{q}$ is weakly structured and relatively flat with respect to $k$, while the wave vector dependence of $\eta_{k}^{q}$ is much more pronounced than that of $\xi_{k}^{q}$. Regardless of coupling strength, the wave vector dependence of $\xi_{k}^{q}$ and $\eta_{k}^{q}$ has a characteristic ``manta ray'' shape, a fact that may be understood through an examination of the real-space structure of the A matrix \cite{zhaoMS}. As revealed in Fig.~\ref{figure1}, the A-matrix parameters tend to have less wave vector dependence for larger $g$. If diagonal coupling strength exceeds that of off-diagonal coupling, the parameters lose most of the $q$ dependence. Thus, the transformation coefficients in Eq.~\ref{variational1}, as approximated by Munn and Silbey \cite{Munn_Silbey2}, represent the A-matrix in Eq.~\ref{variational2} in the limit of strong diagonal coupling. It is also interesting to look into the change of the polaron band structure (especially, band narrowing) as a function of the temperature, which may provide an intuitive understanding of the difference between transport properties obtained here and those by Munn and Silbey using Eq.~\ref{variational1}. Variance of the polaron bandwidth as well as the entire band with respect to temperature has been obtained with our numerically determined A-matrix, and is illustrated in Fig.~\ref{figure2}. It is also indicated in the inset of Fig.~\ref{figure2}(a), the polaron band obtained with the transformation coefficients of Munn and Silbey \cite{Munn_Silbey2} lacks a bimodal structure, and has a much larger bandwidth than that from the current approach. It is indeed counterintuitive to see that the greater
momentum-space variability we find in our transformation coefficients should translate into a narrower polaron band. The explanation lies in that the binding energy is proportional to the average of $|A_k^q|^2$. Compared with results from Ref.~\cite{Munn_Silbey2}, the average binding energy obtained in the current approach is significantly larger and the average Debye-Waller factor significantly smaller, which eventually leads to the narrower band.

To lend further support to the polaron band structure calculated here, it is helpful to introduce the Toyozawa Ansatz \cite{Toyozawa1, Toyozawa2} for zero-temperature band comparisons. As shown in Figs.~\ref{figure2} (b) and (c), good agreement between the current method and the Toyozawa Ansatz has been found. For weak diagonal and off-diagonal coupling as is the case in Fig.~\ref{figure2} (b), the Toyozawa method yields a larger bandwidth, and with increasing $\phi$, the bimodal structure appears and the agreement between the two approaches improves. The concurrence provides legitimacy to the current approach in finding reliable polaron bands, a step that is essential for the transport calculations. Moreover, the band narrowing effect is in good qualitative agreement with those obtained with the Holstein-Peierls model and ab-initio calculations in Ref.~\cite{Hannewald2}. As can be readily seen from Fig.~\ref{figure2} (a), the off-diagonal coupling strength helps boost the bandwidth for temperatures lower than $\omega$, $g^2=0.1$, and $J=0.1$.

To better understand the bandwidth narrowing effect, the parameter matrices $\xi_{k}^{q}$ and $\eta_{k}^{q}$ have been plotted in Fig.~\ref{figure3} for three temperatures, $T = 0.2, 1.3$, and $4.0$ in unit of $\omega$. Other control parameters in Fig.~\ref{figure3} are set to be the same as those for Fig.~\ref{figure2}(c). As temperature increases, the wave vector variations grow for both $\xi_{k}^{q}$ and $\eta_{k}^{q}$, while in Fig.~\ref{figure1}, the variations are reduced as $g$ increases. As explained in Ref.~\cite{zhaoMS}, a larger wave vector variation in $\xi_k^q$ and $\eta_{k}^{q}$ is transformed into a weaker distortion of the energy band, which explains the temperature dependence in Fig.~~\ref{figure2} (a). In contrast to an overall enhancement of polaron band-narrowing by including off-diagonal coupling in Ref.~\cite{Hannewald2}, our model demonstrate the possibility that at low temperatures off-diagonal coupling may increase the bandwidth while at high temperature the effect is reversed. Such an effect can be understood from Eqs.~\ref{epsilontilde}-\ref{selfconsistent3}, where the bandwidth is shown to be closely related to the renormalized transfer integral $\tilde{J}_k$ and the polaron binding energy, which is the third term on the right hand side of  Eq.~\ref{epsilontilde}. At lower temperatures, the binding energy dominates in Eq.~\ref{epsilontilde} in the presence of off-diagonal coupling strength $\phi$, resulting in a bandwidth that increases with $\phi$. At high temperatures, however, the renormalized transfer integral $\tilde{J}_k$ dominates at weak off-diagonal coupling, and the bandwidth derived from the $\tilde{J}_k$ term shrinks with increasing $\phi$. Stronger off-diagonal coupling leads to stronger correlation which helps to form more spatial extended polaron structure\cite{zhaoMS}, which naturally leads to higher coherent phonon-assisted exicton transfer, thus the polaron bandwidth at low temperature for stronger $\phi$ case is lager than the smaller one. However as temperature increases, the inclusion of stronger off-diagonal coupling strength helps to increase the thermal fluctuation of exciton (or polaron) through the renormalized J term in Eq. \ref{epsilontilde}, thus in this case phonon has more significant role in reducing the bandwidth.

\subsection{Transport Properties}

The influence of the diagonal coupling strength $g$ and off-diagonal coupling strength $\phi$ (especially the latter) is substantial in determining momentum space modulations of transformation coefficients $A_k^{q}$, which in turn determine the structure of polaron states and the energy bands. For all calculations here, the number of site is equal to 6 and the hopping integral J is 0.1 in unit of $\omega$ unless specified otherwise. Comparisons between the results obtained here and those from Ref.~\cite{Munn_Silbey3} have been made to illustrate the importance of the A-matrix momentum dependence as well as the straightforward numerical evaluation of transport properties sans additional approximations.

\begin{figure}[tb]
\begin{center}
\includegraphics[scale=0.28]{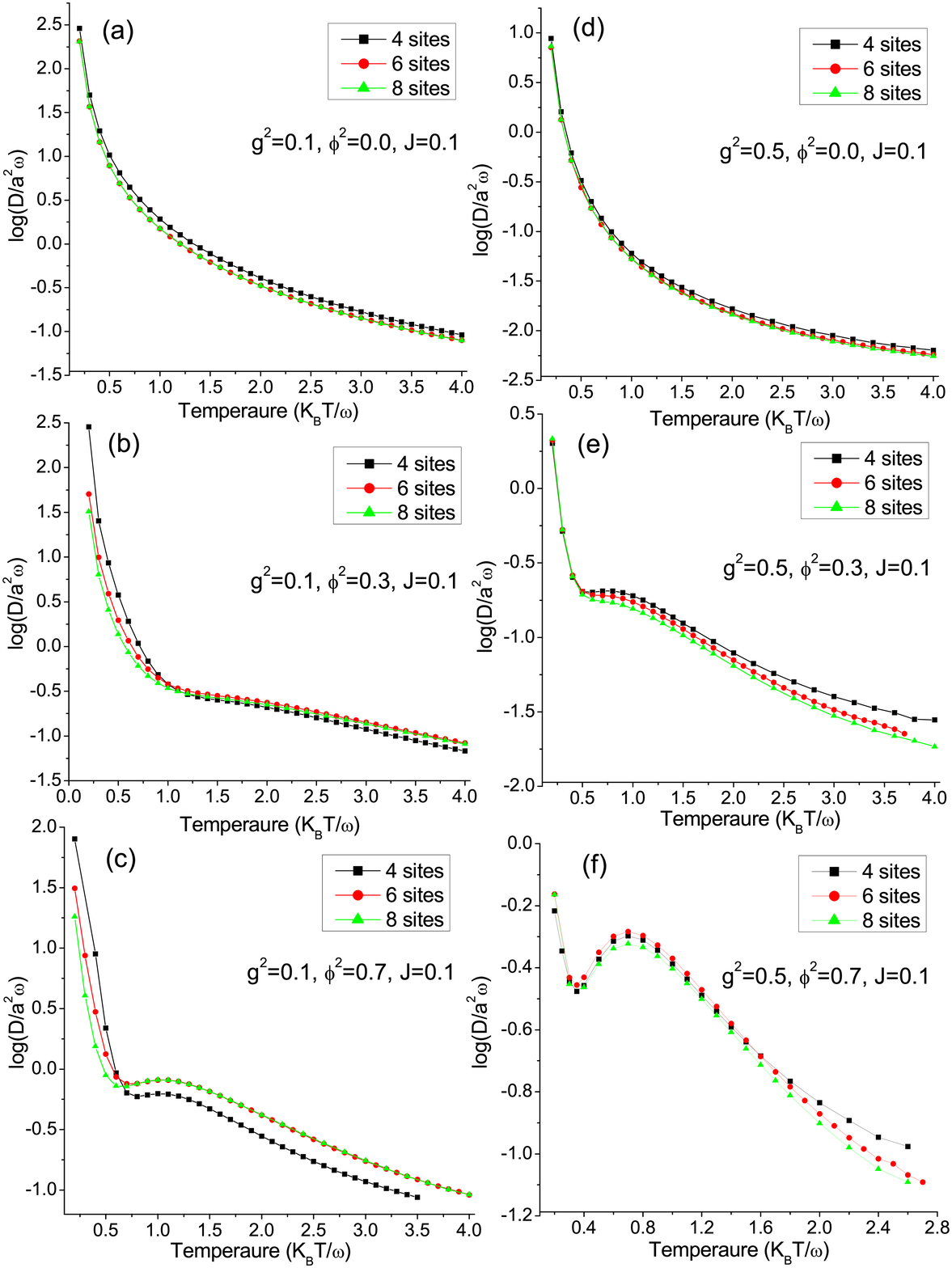}
\end{center}
\caption{Diffusion coefficient D versus scaled temperature $k_{\rm B}{\rm T}/\omega$ for a 1 dimensional chain with 4, 6 and 8 sites, where the diagonal coupling strength $g^2$ = 0.1 for (a)-(c), and $g^2$ = 0.5 for (d)-(f). The off-diagonal coupling strength $\phi^2$ is equal to $0.0$ for (a) and (d), $0.3$ for (b) and (e), and $0.7$ for (c) and (f). Finally, the transfer integral J=0.1 for all cases.}\label{figure4}
\end{figure}

In the calculations of the transport properties, computation time increases dramatically with the number of sites. Our existing computational capabilities allow only calculations for less than 10 sites without overshooting memory requirements. To ensure the robustness of the results obtained with current method, we computed the diffusion coefficient D for 1D lattices of 4, 6 and 8 sites. The results with various lattice sizes and control parameters are displayed in Fig.~\ref{figure4}, indicating that almost identical results are obtained for lattices of 6 and 8 sites in the absence of off-diagonal coupling. Hence, in such a scenario, calculations with 6 sites are expected to give sufficiently reliable values of D at affordable computational cost. As $\phi^2$ increases, however, the number of sites becomes an issue in determining the value of D due to excitation-phonon correlation spans over longer distances. The effect of the lattice size is more prominent in the low temperature regime as shown in Figs.~\ref{figure4}(b) and (c), where the transport properties are dominated by coherent or band-like contributions. At higher temperatures, the hopping contribution eventually dominates the transport and the results become less sensitive to the system size. This can also be understood from polaron band narrowing at high temperatures and a reduced effective transfer integral with increasing temperature. As the diagonal coupling strength $g$ increases, the difference between the results obtained with 4 sites and 6 sites become less obvious than the weak coupling cases. Even better  agreement is reached for $\phi^2$ = 0, and the results are found to be almost identical from 4 to 8 sites as shown in Fig.~\ref{figure4}(d). It is also interesting to spot nonmonotonic changes of the diffusion coefficient (i.e., a ``hump'') as a function of the temperature, in the absence of any size dependence, in the low temperature regime for$g^2 =0.5$ and nonzero $\phi^2$, as shown in Figs.~\ref{figure4} (e) and (f). Such a hump also appears in Fig.~\ref{figure4} (c) for $g^2=0.1$ and $\phi^2=0.7$ despite a slightly visible size dependence of the diffusion coefficient.
For a fixed value of $g^2$, the ``hump'' becomes more pronounced with increasing $\phi^2$, underlying the partial role of off-diagonal coupling.

\begin{figure}[tb]
\begin{center}
\includegraphics[scale=0.33]{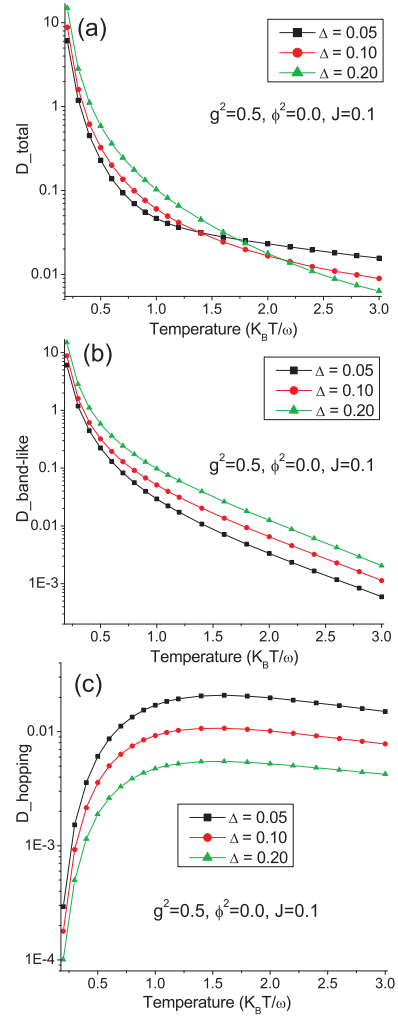}
\end{center}
\caption{Effects of phonon bandwidth $\Delta$ on (a) total diffusion coefficient, (b) band-like contribution and (c) hopping contribution. The other control parameters are $g^2$=0.5, $\phi^2$=0.0 and J=0.1. }\label{figure5}
\end{figure}

As shown in Fig.~\ref{figure5}, increasing the phonon bandwidth $\Delta$ leads to an increase in the band-like contribution to the diffusion coefficient, and a decrease of the relative importance of the hopping contribution. These findings from our numerical calculations for a case of diagonal exciton-phonon coupling are in good agreement with analytical results from Ref.~\cite{Munn_Silbey1}. Qualitatively similar results can be obtained with the inclusion of off-diagonal coupling.

\begin{figure}[tb]
\begin{center}
\includegraphics[scale=0.35]{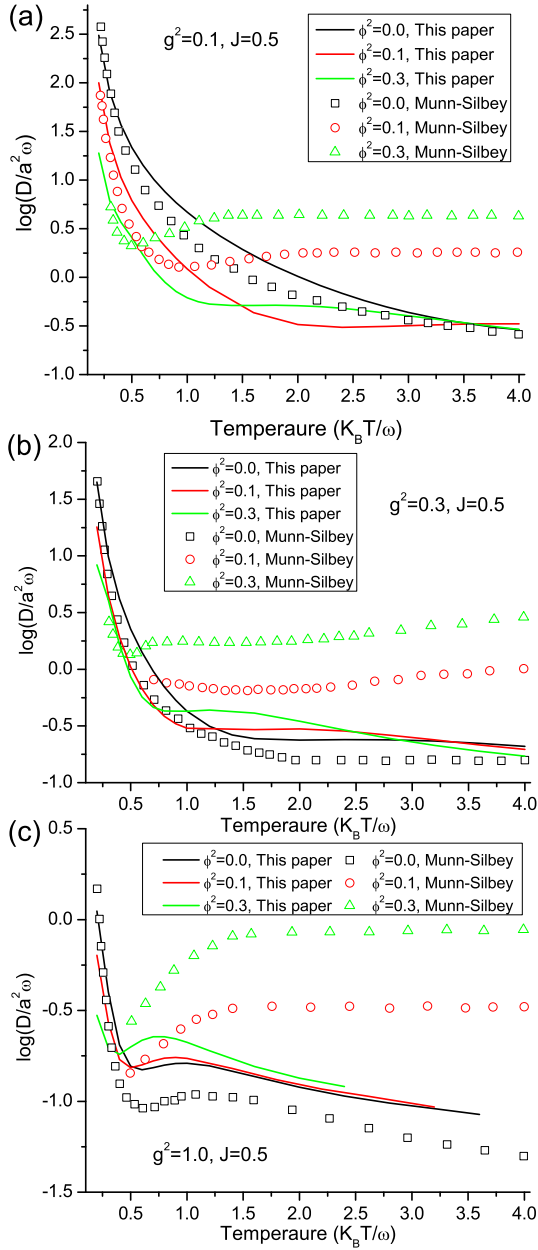}
\end{center}
\caption{Comparisons of the logarithm diffusion coefficient of this work and those from Ref.~\cite{Munn_Silbey3}, where both diagonal and off-diagonal couplings are limited to a moderate value to guarantee the definiteness of our results.}\label{figure6}
\end{figure}

In the original work of Munn and Silbey, the diffusion coefficient as a function of temperature have been studied with structureless A-matrix parameters $\xi$ and $\eta$. The detailed structure for the A-matrix parameters is found to have a substantial impact on the transport properties, as revealed in Fig.~\ref{figure6} by comparing the diffusion coefficients from Ref.~\cite{Munn_Silbey3} with those obtained here. Moreover, we would like to note that in the absence of off-diagonal coupling our results are in good agreement with those from Refs.~\cite{Kenkre1} and \cite{Munn_Silbey4}. The results in this paper can be compared with those obtained using other methods such as a Green's function approach\cite{Fratini} and variational exact diagonalization with a better construction of phonon states\cite{Alvermann}. For example, with an improved numerical technique in Ref.~\cite{Alvermann}, the 1D polaron mass and radius are studied for an expanded parameter regime, where comparisons with our approach are especially convenient. Despite the discrepancies between our results and those of Munn and Silbey in high temperature regime where hopping transport dominates, agreement is found in low temperature regime where band-like transport prevails. This can be understood as follows. As discussed in Ref.~\cite{zhaoMS}, the transformation coefficients obtained self-consistently are shown to have stronger wave vector dependencies as $\phi$ increases. For weak off-diagonal coupling, however, the average values of $\xi_{k}^{q}$ and $\eta_{k}^{q}$ generated by the self-consistency procedure are in reasonable agreement with those by Munn and Silbey \cite{Munn_Silbey2} despite some systematic deviations, which leads to the aforementioned low-temperature agreement. As shown in Figs.~\ref{figure2} and \ref{figure3}, a higher temperature induces a higher amplitude of wave vector variations of the A-matrix parameters together with a much flatter polaron band as compared with its Munn-Silbey counterpart, resulting in a diminished diffusion coefficient. Finally, the diffusion coefficients obtained here in the absence of the off-diagonal coupling are shown to have qualitative agreements with those from Ref.~\cite{Munn_Silbey3}. The differences between the two can be ascribed to a series approximations made in Ref.~\cite{Munn_Silbey3} leading to the calculation of the diffusion coefficient, which in this work are substituted with numerically exact evaluations.

\section{Conclusions}

Within the framework of the Munn-Silbey approach to polaronic transport in organic molecular crystals, diffusion coefficients have been computed with numerical means while taking into account simultaneous diagonal and off-diagonal exciton-phonon coupling for a wide range of temperatures. With the transformation coefficients determined self-consistently without the additional approximations found in Refs.~\cite{Munn_Silbey2,Munn_Silbey3}, calculated transport properties exhibit substantial corrections on those obtained previously thanks to the added momentum-dependence of the A-matrix parameters and numerically exact evaluation of many essential quantities.
The general transport properties obtained are also in good qualitative agreement with those obtained with the Holstein-Peierls model and ab-initio calculations\cite{Hannewald2} demonstrating robustness of the current approach.

Comparisons between this work and the one in Ref.~\cite{Munn_Silbey3} reveal the importance of the electron-phonon correlations and the polaron band structure to transport properties of the extended Holstein molecular crystal model. By treating the diagonal and off-diagonal exciton-phonon coupling on an equal footing in our self-consistency procedure, the current approach is one of the few capable to model realistically the transport phenomenon in organic molecular crystals, for which the effect of off-diagonal coupling often can not be neglected. Off-diagonal coupling, in itself, is a transport mechanism as it is an agent for polaronic localization. Inclusion of off-diagonal coupling in the current approach allows us to see the critical role it plays in determining both polaron structures and transport properties over a wide range of temperatures. Of course, just as any other treatment of transport, the current approach can be improved, especially in the low temperature regime, where the temperature-independent mobility has recently been found. Work in this direction is in progress.

\section*{Acknowledgments}

Support from the Singapore Ministry of Education through the Academic Research Fund (Tier 2) under Project No.~T207B1214 and the Singapore National Research Foundation through the Competitive Research  Programme (CRP) under Project No. NRF-CRP5-2009-04 is gratefully acknowledged.

\appendix

\begin{widetext}

\section{Explicit expressions for the correlation functions}

The expression of the correlation functions $\langle{V_{k'+q,k+q}V_{k,k'}(t)}\rangle$ is mainly composed of three parts as:
\begin{equation}
\langle V_{k_1,k_2}V_{q_1,q_2}(t)\rangle =X_{k_1,k_2;q_1,q_2}+Y_{k_1,k_2;q_1,q_2}+\Lambda_{k_1,k_2;q_1,q_2}.
\end{equation}
First of all, the expression of $X_{k_1,k_2;q_1,q_2}$ is given by:
\begin{eqnarray}\label{termAA}
&&X_{k_1,k_2;q_1,q_2} \nonumber \\
&=&\underset{k_3,k_4}{\sum }[J_{k_3}J_{k_4}\langle T_{k_3,k_1;k_3,k_2}T_{k_4,q_1;k_4,q_2}\rangle \nonumber \\
&-&\frac{1}{N}\underset{Q_2}{\sum }J_{k_3}\omega _{Q_2}f_{-k_4}^{Q_2}\cdot(A_{-q_1}^{-Q_2}\langle T_{k_3,k_1;k_3,k_2}T_{k_4+Q_2,q_1;k_4,q_2-Q_2}\rangle+A_{-q_1}^{Q_2\ast }\langle T_{k_3,k_1;k_3,k_2}T_{k_4+Q_2,q_1+Q_2;k_4,q_2}\rangle ) \nonumber \\
&-&\frac{1}{N}\underset{Q_1}{\sum }J_{k_4}\omega _{Q_1}f_{-k_3}^{Q_1}\cdot(A_{-k_1}^{-Q_1}\langle T_{k_3+Q_1,k_1;k_3,k_2-Q_1}T_{k_4,q_1;k_4,q_2}\rangle+A_{-k_1}^{Q_1\ast }\langle T_{k_3+Q_1,k_1+Q_1;k_3,k_2}T_{k_4,q_1;k_4,q_2}\rangle ) \nonumber \\
&+&\frac{1}{N^{2}}\underset{Q_1,Q_2}{\sum }\omega _{Q_1}\omega_{Q_2}f_{-k_3}^{Q_1}f_{-k_4}^{Q_2} \nonumber \\
&\times&(A_{-k_1}^{-Q_1}A_{-q_1}^{-Q_2}\langle T_{k_3+Q_1,k_1;k_3,k_2-Q_1}T_{k_4+Q_2,q_1;k_4,q_2-Q_2}\rangle+A_{-k_1}^{-Q_1}A_{-q_1}^{Q_2\ast }\langle T_{k_3+Q_1,k_1;k_3,k_2-Q_1}T_{k_4+Q_2,q_1+Q_2;k_4,q_2}\rangle \nonumber \\
&+&A_{-k_1}^{Q_1\ast }A_{-q_1}^{-Q_2}\langle T_{k_3+Q_1,k_1+Q_1;k_3,k_2}T_{k_4+Q_2,q_1;k_4,q_2-Q_2}\rangle +A_{-k_1}^{Q_1\ast }A_{-q_1}^{Q_2\ast }\langle T_{k_3+Q_1,k_1+Q_1;k_3,k_2}T_{k_4+Q_2,q_1+Q_2;k_4,q_2}\rangle)],
\end{eqnarray}
where
\begin{equation}
\langle T_{k_1,k_2;k_3,k_4}T_{q_1,q_2;q_3,q_4}(t)\rangle=\langle\theta _{k_1,k_2}^{\dag }\theta _{k_3,k_4}\theta _{q_1,q_2}^{\dag}(t)\theta _{q_3,q_4}(t)\rangle -\langle \theta _{k_1,k_2}^{\dag }\theta_{k_3,k_4}\rangle\langle \theta _{q_1,q_2}^{\dag }\theta _{q_3,q_4}\rangle,
\end{equation}
and the expression for $\langle\theta _{k_1,k_2}^{\dag }\theta _{k_3,k_4}\theta _{q_1,q_2}^{\dag}(t)\theta _{q_3,q_4}(t)\rangle$ and $\langle \theta _{k_1,k_2}^{\dag }\theta_{k_3,k_4}\rangle\langle \theta _{q_1,q_2}^{\dag }\theta _{q_3,q_4}\rangle$ has been given by Eq.~\ref{fourtheta} and Eq.~\ref{twotheta}, respectively. Here and onwards, $N$ represents the total number of sites used in the transport calculation.\\

Now we turn to the expression of $Y_{k_1,k_2;q_1,q_2}$, which is written as:
\begin{eqnarray}
&&Y_{k_1,k_2;q_1,q_2}\nonumber\\
&=&\underset{k_3,k_4}{\sum }\{\frac{1}{\sqrt{N}}[\underset{Q}{\sum }J_{k_3}\omega_{Q}f_{-k_4}^{Q}\langle Z_{k_1,k_3;k_3,k_2}\psi _{Q}(t)\rangle \langle T_{k_3,k_1;k_3,k_2}T_{k_4+Q_,q_1;k_4,q_2}(t)\rangle-\frac{1}{N}\underset{Q_1,Q_2}{\sum }\omega _{Q_1}\omega_{Q_2}f_{-k_3}^{Q_1}f_{-k_4}^{Q_2}\nonumber\\
&\times&(A_{-k_1}^{-Q_1}\langle Z_{k_1,k_3+Q_1;k_3,k_2-Q_1}\psi _{Q_2}(t)\rangle\langle T_{k_3+Q_1,k_1;k_3,k_2-Q_1}T_{k_4+Q_2,q_1;k_4,q_2}(t)\rangle \nonumber\\
&+&A_{-k_1}^{Q_1\ast }\langle Z_{k_1+Q_1,k_3+Q_1;k_3,k_2}\psi _{Q_2}(t)\rangle\langle T_{k_3+Q_1,k_1+Q_1;k_3,k_2}T_{k_4+Q_2,q_1;k_4,q_2}(t)\rangle)] \nonumber\\
&+&\frac{1}{\sqrt{N}}[\underset{Q_1}{\sum }J_{k_4}\omega _{Q_1}f_{-k_3}^{Q_1}\cdot\langle \psi _{Q_1}Z_{q_1,k_4;k_4,q_2}(t)\rangle \langle T_{k_3+Q_1,k_1;k_3,k_2}T_{k_4,q_1;k_4,q_2}(t)\rangle -\frac{1}{N}\underset{Q_1,Q_2}{\sum }\omega_{Q_1}\omega_{Q_2}f_{-k_3}^{Q_1}f_{-k_4}^{Q_2}\nonumber\\
&\times&(A_{-q_1}^{-Q_2}\langle \psi _{Q_1}Z_{q_1,k_4+Q_2;k_4,q_2-Q_2}(t)\rangle\langle T_{k_3+Q_1,k_1;k_3,k_2}T_{k_4+Q_2,q_1;k_4,q_2-Q_2}(t)\rangle\nonumber\\
&+&A_{-q_1}^{Q_2\ast }\langle \psi _{Q_1}Z_{q_1+Q_2,k_4+Q_2;k_4,q_2}(t)\rangle\langle T_{k_3+Q_1,k_1;k_3,k_2}T_{k_4+Q_2,q_1+Q_2;k_4,q_2}(t)\rangle)]\},
\end{eqnarray}
where
\begin{equation}
\langle C_{k_1,k_2;k_3,k_4}\psi_{q}(t)\rangle =\frac{1}{\sqrt{N}}\{A_{-k_2}^{k_1-k_2}[ne^{-i\omega t}-(1+n)e^{i\omega t}]\delta _{-k_1+k_2,q}-A_{-k_4}^{k_3-k_4}[ne^{-i\omega t}-(n+1)e^{i\omega t}]\delta _{-k_3+k_4,q}\},
\end{equation}
and
\begin{equation}
\langle\psi_{q}Z_{k_1,k_2;k_3,k_4}(t)\rangle = \frac{1}{\sqrt{N}}\{A_{-k_2}^{k_1-k_2}[(n+1)e^{i\omega t}-ne^{-i\omega t}]\delta_{q,-k_1+k_2}-A_{-k_4}^{k_3-k_4}[(n+1)e^{i\omega t}-ne^{-i\omega t}]\delta _{q,-k_3+k_4}\},\\
\end{equation}
here $n$ is the Bose-Einstein distribution, with $2n+1= {\rm coth}(\frac{1}{2}\beta\hbar\omega)$.\\

Finally we give the expression of $\Lambda_{k_1,k_2;q_1,q_2}$:

\begin{eqnarray}
&&\Lambda_{k_1,k_2;q_1,q_2} \nonumber\\
&=&\underset{k_3,k_4}{\sum }\frac{1}{N}\{\underset{Q_1Q_2}{\sum }\omega_{Q_1}\omega _{Q_2}f_{-k_3}^{Q_1}f_{-k_4}^{Q_2}\nonumber \\
&\times&\{\frac{1}{N}(-A_{-k_3-Q_1}^{k_1-k_3-Q_1}A_{-k_4-Q_2}^{q_1-k_4-Q_2}\delta _{k_1-k_3-Q_1,-Q_2}\delta_{-Q_1,q_1-k_4-Q_2}+A_{-k_3-Q_1}^{k_1-k_3-Q_1}A_{-q_2}^{k_4-q_2}\delta_{k_1-k_3-Q_1,-Q_2}\delta _{Q_1,q_2-k_4} \nonumber \\
&+&A_{-k_2}^{k_3-k_2}A_{-k_4-Q_2}^{q_1-k_4-Q_2}\delta _{k_2-k_3,Q_2}\delta_{-Q_1,q_1-k_4-Q_2}-A_{-k_2}^{k_3-k_2}A_{-q_2}^{k_4-q_2}\delta _{k_2-k_3,Q_2}\delta_{Q_1,q_2-k_4})\cdot[ne^{-i\omega t}-(n+1)e^{i\omega t}]^{2} \nonumber\\
&+&\langle \psi _{Q_1}\psi _{-Q_2}(t)\rangle\delta _{Q_1,-Q_2}\}\cdot \langle T_{k_3+Q_1,k_1;k_3,k_2}T_{k_4+Q_2,q_1;k_4,q_2}\rangle\}
\end{eqnarray}

\end{widetext}

\end{document}